\documentclass{aa}

\usepackage{graphicx}
\usepackage{txfonts}
\usepackage{tikz}
\usepackage{xcolor}
\definecolor{darkgreen}{rgb}{0.0,0.5,0.0}
\usepackage{hyperref}
\hypersetup{
    colorlinks=true,
    citecolor=darkgreen,
    linkcolor=blue,
    filecolor=magenta,
    urlcolor=cyan,
    pdftitle={Shocks at RB edges}
    }
\usepackage[normalem]{ulem}

\makeatletter
\let\AA@orig@oddhead\@oddhead
\def\@evenhead{}

\makeatother

\newcommand{\eg}{\emph{e.g.,} }

\newcommand{\be}{\begin{equation}}
\newcommand{\ee}{\end{equation}}
\newcommand{\bea}{\begin{equation*}}
\newcommand{\eea}{\end{equation*}}
\newcommand{\beqr}{\begin{eqnarray} \nonumber}
\newcommand{\eeqr}{\end{eqnarray}}
\newcommand{\beqrb}{\begin{eqnarray}}
\newcommand{\eeqrb}{\nonumber \end{eqnarray}}

\newcommand{\coma}{\mbox{ ,}}

\newcommand{\cm}{\mbox{ cm}}

\newcommand{\Myr}{\mbox{ Myr}}

\newcommand{\MHz}{\mbox{ MHz}}
\newcommand{\GHz}{\mbox{ GHz}}

\newcommand{\eV}{\mbox{ eV}}
\newcommand{\keV}{\mbox{ keV}}

\newcommand{\GeV}{\mbox{ GeV}}

\newcommand{\muG}{\mbox{ $\mu$G}}

\newcommand{\gama}{$\gamma$}

\newcommand{\mynewcommand}[2]{\ifdefined #1 \else \newcommand{#1}{#2} \fi}

\mynewcommand{\apj}{ApJ}     
\mynewcommand{\apjl}{ApJL}     
\mynewcommand{\apjs}{ApJS}    
\mynewcommand{\aap}{A\&A}    
\mynewcommand{\nat}{Nature}  

\newcommand{\dgr}{^{\circ}}

\def\myfig#1{Figures/#1}
\newcommand{\Mach}{\mathcal{M}}

\newcommand*{\ShowComments}{}
\ifdefined\ShowComments
    \newcommand{\Uri}[1]{{\textcolor{orange}{\tt[#1]}}}
    \newcommand{\Arka}[1]{{\textcolor{blue}{\tt[#1]}}}
    \newcommand{\ArkaCorrectionB}[2]{\sout{#1}\textcolor{cyan}{#2}}
\else
    \newcommand{\Uri}[1]{}
    \newcommand{\Arka}[1]{}
    \newcommand{\ArkaCorrectionB}[2]{}
\fi

\begin{document}

   \title{Mach $>3$ shocks at the tips of both eROSITA bubbles}
   \titlerunning{Mach $>3$ shocks at the tips of both eROSITA bubbles}

   \author{
       Uri Keshet\thanks{\email{keshet.uri@gmail.com}}
       \and
       Arka Ghosh
   }

   \institute{Physics Department, Ben-Gurion University of the Negev, POB 653, Be'er-Sheva 84105, Israel}

    \date{\today}


  \abstract
   {eROSITA substantiated earlier indications that Loop-I is the northern part of an extended bipolar Galactic-bubble structure, but the southern bubble was not established in nonthermal emission and the shock strength was not robustly measured in either bubble.}
   {After using eROSITA data to map the bubble edges, we analyzed edge-adjacent radio and $\gamma$-ray data to remove foregrounds, test if the southern bubble can be detected in nonthermal emission, and measure the corresponding high-latitude spectra of both bubbles. }
   {Data were stacked parallel to the eROSITA bubble edges traced by an edge detector, in the same method used previously to pick up weak signals in the smaller, nested Fermi bubbles; the detected brightness jumps were then used to measure the spectrum. }
   {We detect ($>5\sigma$) both bubble tips in both radio and $\gamma$-rays, and find a radio spectrum corresponding to high, Mach $3$--$5$ shocks. The southern bubble is fainter, by $\sim$an order of magnitude in radio, its edge propagating into a medium roughly half as dense. }
   {The results indicate that these eROSITA bubbles are older, evolved counterparts of the Fermi bubbles, arising from an earlier collimated high-energy outburst from the Galactic center. }


   \maketitle
   \thispagestyle{empty}
%

\section{Introduction}

A pair of extended bipolar bubbles, originating from the Milky-Way Galactic center and reaching high, $|b|\sim 80^\circ$ latitudes, was indicated by radio \citep{Sofue77} and X-rays \citep{Sofue94}, in particular \textit{ROSAT} observations \citep{Sofue00}, and more recently established by \textit{eROSITA} data \citep{PredehlEtAl20}.
The northern bubble or part thereof is also known as Loop-I or the north polar spur (NPS).
These bubbles, henceforth designated the RBs, are similar but more extended than their Fermi bubble \citep[][henceforth FBs]{SuEtAl10} counterparts, which are nested \citep[not only in projection; see][]{GhoshEtAl26}, within the RBs and reach only $\vert b\vert\sim 50^\circ$. For a review, see \citet{Sarkar24}.

The edges of the RBs are thought to be outgoing (forward) shocks, heating the gas, magnetizing the plasma, and accelerating charged particles to high energies, in resemblance of the FB edges \citep[distinguished most clearly from alternative interpretations by their X-ray shells;][]{Keshetgurwich18}.
Shock Mach numbers inferred from the electron temperature jump across the NPS span the wide range $\Mach\sim 1.4\text{--2.7}$ \citep{AkitaEtAl18, LaRoccaEtAl20, YamamotoEtAl22, Sarkar24}, under the assumption that the RBs are sufficiently old for electron--ion thermalization in the measured downstream region.
Note that this is not the case for the few-Myr old FBs, where the electron temperature jumps based on thermal X-rays \citep{Keshetgurwich18} or \ion{O}{VII} and \ion{O}{VIII} line emission \citep{MillerBregman2015, MillerBregman2016} correspond to Mach numbers smaller than inferred from the hard spectra of radio \citep{KeshetEtAl24} and \gama-ray \citep{Keshetgurwich17} emission from the bubble edge, bubble-integrated microwaves \citep{Dobler12, PlanckHaze13} and \gama-rays \citep{SuEtAl10}, and the nearly uniform \gama-ray spectrum along the shock front \citep{Keshetgurwich17}.

The shock strength can be inferred robustly from low-frequency radio emission, below any cooling break, adopting standard Fermi acceleration in a leptonic model, as implied by observations (see \S\ref{sec:Discussion}), whereby the photon spectral index
\begin{equation}
\alpha\equiv -\frac{d\ln(I_\nu)}{d\log(\nu)} \simeq \frac{\Mach^2+3}{2(\Mach^2-1)}\coma
\label{eq:alpha}
\end{equation}
where $I_\nu$ is the specific brightness at frequency $\nu$.
However, spectral indices attributed to the RBs, including
$\alpha=0.3$--$1$ \citep{ReichReich88}, 
$\alpha=0.74\pm0.08$ \citep{Borka07},
$\alpha=0.55$--$0.65$ \citep{Guzman11},
$\alpha\simeq 0.4$--$0.7$ \citep{IwashitaEtAl23},
and
$\alpha\simeq 0.55$--$0.60$ \citep{MouEtAl23}, are not consistent with each other.
The interpretation of the radio--microwave spectrum, \eg $\alpha=1.07_{-0.09}^{+0.13}$ between $0.4$ and $23\GHz$ \citep{DaviesEtAl06}, is sensitive to the spatially-integrated location of the anticipated cooling break energy, as demonstrated in the FBs \citep{KeshetEtAl24}.
The \gama-ray signal, detected until now only in the north RB, is in general soft, weak, and unresolved \citep[\eg][]{SuEtAl10}; its interpretation \citep[\eg][]{MouEtAl23} is sensitive to the poorly constrained Galactic radiation fields at the emitting region.

Such analyses all focus on the northern RB, and in general do not remove foregrounds and backgrounds (henceforth foreground), nor isolate the near-edge spectrum, so they are sensitive to the region examined.
Numerical simulations have not, in general, reproduced observations, giving for example Mach numbers that are very low \citep[$\Mach\lesssim1.5$;][]{ZhangEtAl25} or very high \citep[$\Mach\sim 10$;][]{YangEtAl22}. 
The FB edges are often modelled as weak forward shocks or alternative discontinuities or RB-related structures \citep[\eg][and references therein]{TsengEtAl24, Sarkar24}, also inconsistent with recent observations.

\tikzset{
  north arrow/.pic={
    \draw[fill=white,draw=white,line width=0.4pt]
      (0,-1.5) -- (-1,0) -- (0,1.5) -- (1,0) -- cycle;     
    \draw[fill=black,draw=black,line width=0.4pt]
      (0,0.6) -- (-0.25,0) -- (0.25,0) -- cycle;           
    \draw[fill=white,draw=black,line width=0.4pt]
      (0,-0.6) -- (-0.25,0) -- (0.25,0) -- cycle;          
    \node[above,font=\sffamily\tiny] at (0,0.5) {N};       
  }
}

\begin{figure*}[h]
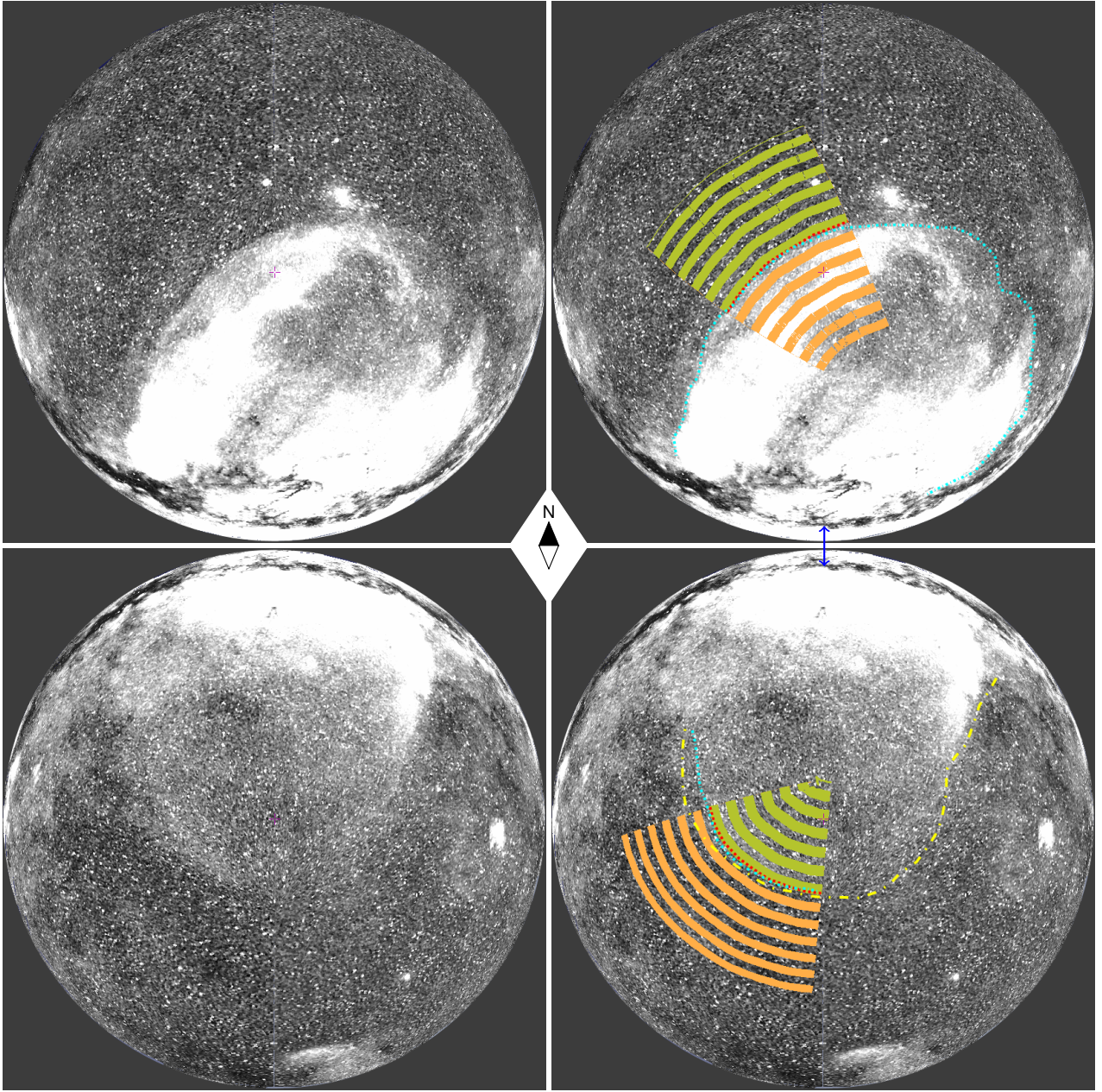

        \begin{tikzpicture}
            \draw (0, 0) node[inner sep=0]
            {
            \centerline{\includegraphics[width=0.45\linewidth,trim={9.3cm 0.7cm 9.3cm 0.7cm},clip]{\myfig{SaveN0c.png}}
                        \includegraphics[width=0.45\linewidth,trim={9.3cm 0.7cm 9.3cm 0.7cm},clip]{\myfig{SaveN3c2.png}}}
            };
            \draw (0, -8.4) node[inner sep=0]
            {
            \centerline{\includegraphics[width=0.45\linewidth,trim={9.3cm 0.7cm 9.3cm 0.7cm},clip]{\myfig{SaveS0c.png}}
                        \includegraphics[width=0.45\linewidth,trim={9.3cm 0.7cm 9.3cm 0.7cm},clip]{\myfig{SaveS3c2.png}}}
            };
            \draw[<->,blue,thick]  (4.20,-3.9)   -- (4.20,-4.52);
            \pic [scale=0.6] at (0,-4.2) {north arrow};
       \end{tikzpicture}
       \caption{\label{fig:eROSITA}
       eROSITA $0.6$--$1\keV$ spherical-projection images of north (top row) and south (bottom) RBs, centered on Galactic coordinates $(l,b)=(0,\pm70\dgr)$.
       Superimposed on right panels are detected edges based on short (dot-dashed cyan curve) and long (dot-long-dashed yellow) gradients, the edge section used for binning (dotted red), its equal-angle distance-transform bins (shaded regions), and the Galactic center (blue arrows).
    }
\end{figure*}

As the RB edges are nicely traced by the eROSITA sky map, it is possible to stack multi-messenger data in bins parallel to the edge, as a function of the oriented distance $\psi$ from the shock, and examine the sudden change in brightness near the $\psi=0$ edge due to the shock.
This method raises the sensitivity to weak signals, as demonstrated by facilitating the detection of the embedded high-latitude FB shells in X-rays \citep{Keshetgurwich18} and \gama-rays \citep{Keshetgurwich17}, and in both polarized \citep{Keshet25PolFB} and non-polarized \citep{KeshetEtAl24} synchrotron and dust emission.
We apply this method to radio and \gama-ray data around the RBs, to examine if their edges can be detected at high latitudes and if their spectra can be measured.

\section{Stacking data along the RB edge}
\label{sec:Method}

The eROSITA all-sky map \citep{PredehlEtAl20} outlines the RB edges most clearly in its $0.6$--$1.0\keV$ channel.
Figure \ref{fig:eROSITA} reproduces this channel separately in spherical projections of the north (top row) and south (bottom) Galactic hemispheres, and demonstrates (right panels) a first-order directional Gaussian-derivative edge detector \citep{canny1986computational} of $\sim5\dgr$ smoothing scale (dot-dashed cyan). 
In the south, such small-scale gradients do not capture the full extent of the bubble, so the detected edge is also shown for extended smoothing ($\sim15\dgr$; dot-long-dashed yellow).
The edges in both hemispheres are clearest east of the bubble tip, so our nominal $\psi=0$ edge (dotted red) is based on the short-scale gradients in the eastern sectors shown in the figure.
Bins parallel to the nominal edge outside ($\psi>0$) and inside ($\psi<0$) each bubble are highlighted as superimposed alternating shaded regions. Our results are not sensitive to small variations in edge or bin definitions, as demonstrated in \S\ref{sec:Results}.

The detected RB edges are analyzed in the same method applied previously to the FB edges \citep[][and references therein]{Keshet25PolFB}, as follows.
Focusing on the vicinity of the RB edges, we use the surrounding bins to stack radio and \gama-ray data in different channels, as well as the reference eROSITA data, in order to measure the respective brightness profiles across each bubble edge as a function of $\psi$, and to use the brightness jump near $\psi=0$ to infer the bubble spectra in its near downstream.
The choice of data for such stacking is restricted to sky maps that cover the RBs in both hemispheres and show sufficient resolution with accurate large-scale zero-level and background stability.

The low-frequency radio data thus chosen include the $45\MHz$ \citep{AlvarezEtAl97, MaedaEtAl99}, $150\MHz$ \citep{LandeckerWielebinski70}, $408\MHz$ \citep{Haslam82}, and $1.4\GHz$ all-sky maps.
For the latter, in the south we use the CHIPASS continuum map \citep{CalabrettaEtAl14}. Due to the partial coverage of CHIPASS at high declinations upstream of the north RB, in the northern hemisphere we use, instead, the Stockert--Villa Elisa map \citep{ReichEtAl01}.

For the \gama-ray data, we examine four channels logarithmically spaced in the $0.1$--$10\GeV$ range (labelled channels 1--4, from low to high energy), using the archival Pass-8 LAT data from the Fermi Science Support Center (FSSC)\footnote{See \url{http://fermi.gsfc.nasa.gov/ssc}}, and the Fermi Science Tools (version \texttt{v10r0p5}).
Weekly all-sky files spanning weeks $9$ through $789$ for a total of $781$ weeks ($\sim15$ years) are used, with ULTRACLEANVETO class photon events.
We apply a $90^\circ$ zenith angle cut to avoid CR-generated $\gamma$-rays from the Earth's atmospheric limb, according to
FSSC recommendations, and select good time intervals using the recommended expression \texttt{(DATA\_QUAL==1) and (LAT\_CONFIG==1)}.
Point-source contamination is minimized by masking pixels within the $68\%$ total-event containment area of each point source in the LAT fourth source catalog \cite[4FGL;][]{AbdollahiEtAl20}, but we cap the masking radius at a $2\dgr$ maximum to avoid bins of insufficient pixels at low energy.

A non-Euclidean, signed distance transform of each map on the plane of the sky yields the oriented angular distance $\psi$ of each pixel from the bubble edge. The resulting $I_\nu(\psi)$ specific-brightness profile is binned at fixed $\psi$ intervals, nominally of $2\dgr$ width, and a foreground model $F_\nu(\psi)$ is subtracted.
Focusing on short, $|\psi|\lesssim 20^\circ$ distances from the edge, a constant foreground provides a reasonable approximation at the relevant high latitudes, although we also consider a linear $F_\nu(\psi)$ fit.
Nominally, we define upstream $\Psi_u$ and downstream $\Psi_d$ regions $6\dgr$--$10\dgr$ from the edge, to minimize the effect of edge localization errors, and adopt a fixed $F_\nu=\langle I_\nu(\Psi_u)\rangle$ averaged upstream; the results depend weakly on these choices, as demonstrated in \S\ref{sec:Results}.

A sudden increase in $I'_\nu(\psi)$ around $\psi\simeq 0$ would suggest emission associated with the RBs, of brightness $I_\nu(\psi<0)-F_\nu$.
The detection significance is inferred from the $\mbox{TS} = \chi^2_- - \chi^2_+$ test, which compares the $\mathsf{n}$ degree-of-freedom $\chi^2$ fit values obtained before ($-$ subscripts) and after ($+$ subscript) adding a simple, nominally linear $\Delta I(\psi<0)$ excess to the foreground $F_\nu(\psi)$; TS then approximately follows \citep{Wilks1938} a chi-squared distribution $\chi_\mathsf{n}^2$ of order $\mathsf{n}\equiv \mathsf{n}_+-\mathsf{n}_-$.
A putative opposite, sudden decrease, in $I'_\nu(\psi\simeq0)$ would similarly indicate the presence of an edge, but not of excess bubble emission.

\section{Results}
\label{sec:Results}

\begin{figure}[h!]
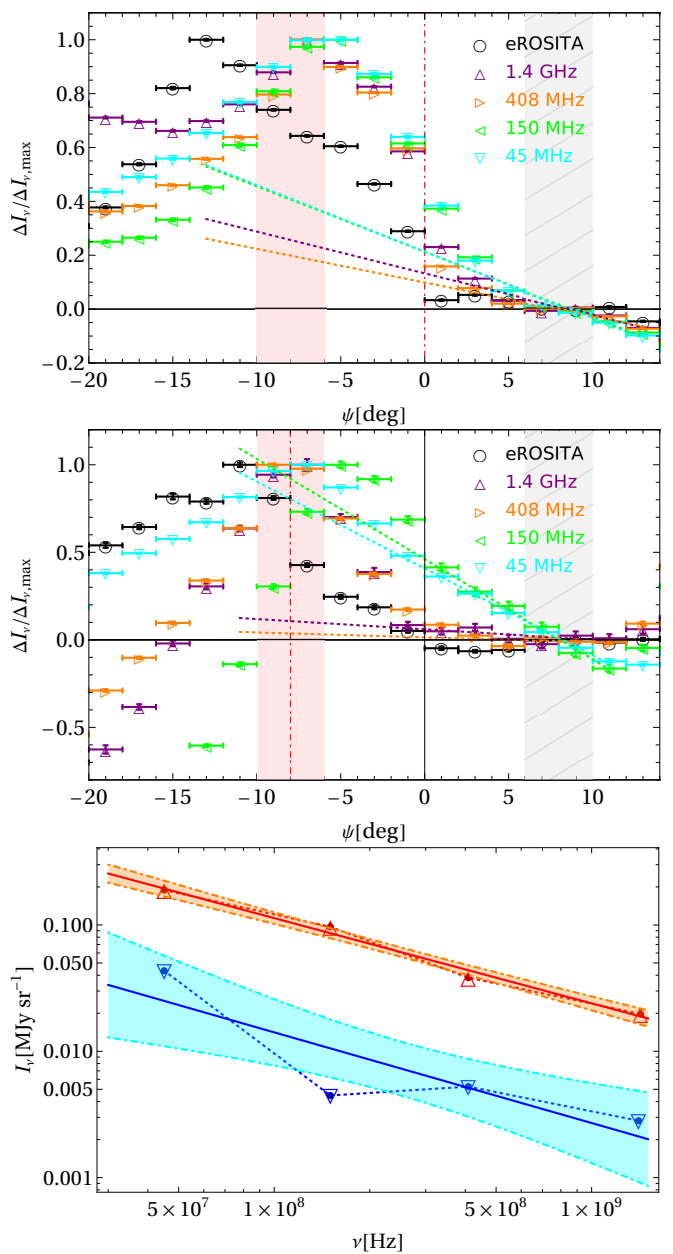

    \centerline{\includegraphics[width=0.95\linewidth,trim={0cm 0cm 0cm 0cm},clip]{\myfig{ProfN1c2.eps}}}
    \centerline{\includegraphics[width=0.95\linewidth,trim={0cm 0cm 0cm 0cm},clip]{\myfig{ProfS1b.eps}}}
    \centerline{\includegraphics[width=0.95\linewidth,trim={0cm 0cm 0cm 0cm},clip]{\myfig{RadSpect1.eps}}}
        \caption{\label{fig:Radio}
       Brightness profiles across north (top panel) and south (middle) RB edges, and the inferred radio spectra in the near downstream (bottom), in the nominal analysis.
       The X-rays (circles) and radio (triangles of different colors and orientations designating different bands, see legend) profiles are shown (including error bars), along with nominal upstream ($\psi>0$, hatched) and downstream regions (shaded) and linear fits to the upstream radio brightness (dotted lines); edge detectors pick up the steepest X-ray gradient (vertical dot-dashed red lines).
       The spectrum is inferred from the excess brightness $I_\nu-F_\nu$ in the four radio channels (error bars with dotted lines to guide the eye) in the north (up red triangles) and south (down blue triangles); best linear fits are shown (solid curves with shaded $1\sigma$ uncertainty range).
    }
\end{figure}

\begin{figure}[h!]
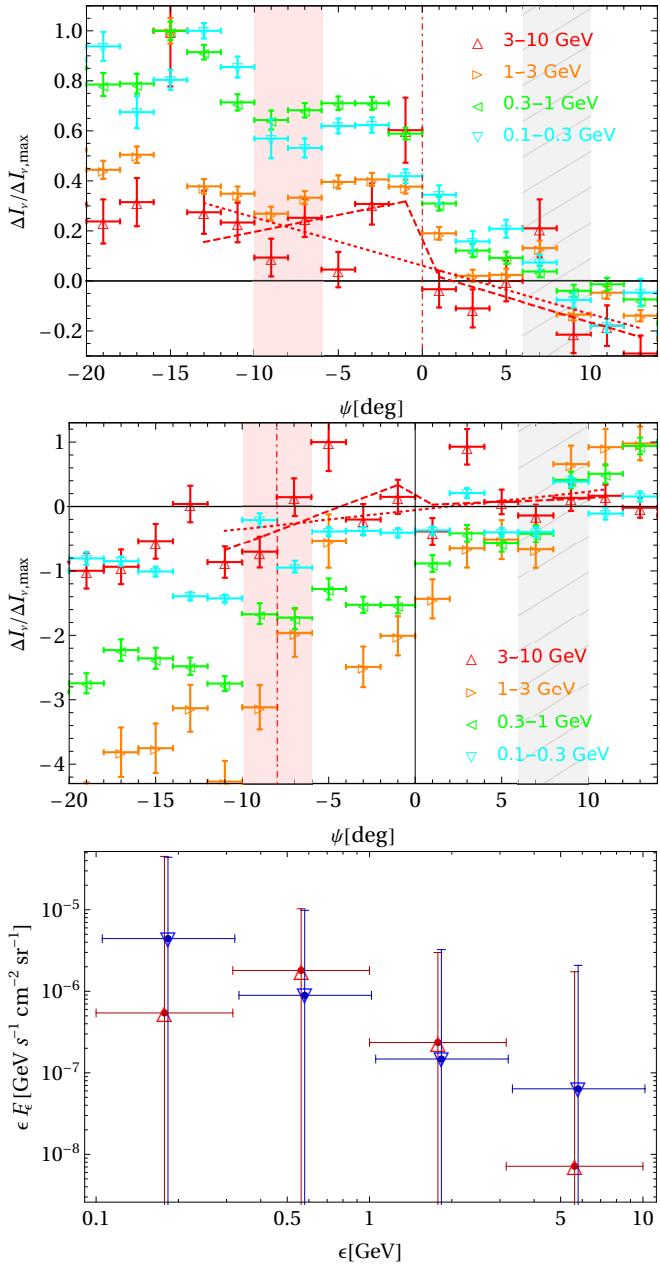

    \centerline{\includegraphics[width=0.95\linewidth,trim={0cm 0cm 0cm 0cm},clip]{\myfig{ProfN1G2c.eps}}}
    \centerline{\includegraphics[width=0.95\linewidth,trim={0cm 0cm 0cm 0cm},clip]{\myfig{ProfS1G2.eps}}}
    \centerline{\includegraphics[width=0.95\linewidth,trim={0cm 0cm 0cm 0cm},clip]{\myfig{GammaSpect3.eps}}}
    \caption{\label{fig:GammaRay}
       Same as Fig.~\ref{fig:Radio} for the four \gama-ray channels (see legends).
       Linear (dotted red lines) and piecewise linear (dashed) fits are demonstrated for the $I_\nu(\psi)$ profiles in the high, $3$--$10\GeV$ energy channel.
       In the bottom panel, down triangles depicting the spectrum of the south bubbles are slightly shifted horizontally for visibility.
    }
\end{figure}

Fairly sharp breaks in the binned $I_\nu(\psi)$ profiles are observed in all inspected X-ray, radio (see Fig.~\ref{fig:Radio}), and \gama-ray (Fig.~\ref{fig:GammaRay}) channels, across both north (top panels) and south (middle panels) RB edges.
The precise shock location $\psi=0$ is matched to the initial rise in X-ray profile above the foreground, which happens to coincide with the steepest gradient picked up by the edge detector (dot-dashed vertical red lines) in the north, but is shifted upstream by $8\dgr$ in the south.
Stacking the data over the extended sectors reduces the noise (error bars designate standard deviations), so the edges are detected at a high ($>5\sigma$) confidence level in all radio channels, albeit not in all \gama-ray channels discussed below.
The stacked $\psi\simeq0$ transitions remain pronounced despite imprecisions in edge localization, possible edge substructure, and foreground variations.

Deprojection of the bubble's line-of-sight projected emission is challenged by the non-trivial RB geometry and unrelated foreground features within the large radius of curvature, so the brightness excess in each channel is crudely estimated directly from the projected profiles after foreground subtraction, as outlined in \S\ref{sec:Method}.
The top and middle panels of Fig.~\ref{fig:Radio} highlight the nominal upstream and downstream regions (shaded vertical bands) and illustrate two methods of foreground estimation: either as a constant $F_\nu$ derived by averaging over the upstream (hatched, $\psi>0$ shaded) region, or as a linear $F_\nu(\psi)$ fit (dotted lines) to the upstream brightness.
The former is used to compute the nominal $\Delta I_\nu(\psi)\equiv I_\nu(\psi)-F_\nu$ excess (ordinate), which approximates the foreground on these scales as constant and is thus independent of overall additive (zero-point), albeit not multiplicative (scale/gain), errors in the raw data;
the ordinate is further normalized by its maximal value in each channel for visual purposes, to simultaneously depict multiple channels.

The inferred downstream excess $\Delta I_\nu$ is then used to measure the RB spectrum averaged along its edge as a function of $\nu$, as shown in the bottom panels of Figs.~\ref{fig:Radio} and ~\ref{fig:GammaRay} for the four radio and four \gama-ray channels, respectively.
Consider first the radio spectrum, which can be measured at a higher precision.
The nominal analysis yields a spectral index $\alpha=0.67\pm0.05$ in the northern RB, corresponding by Eq.~\eqref{eq:alpha} to a Mach $\Mach=3.5_{-0.4}^{+0.7}$ shock.
The fainter excess of the southern bubble yields a less certain $\alpha=0.72\pm0.29$, corresponding to $\Mach=3.2_{-1.0}^{+1.7}$.
The southern $150\MHz$ excess, while significant due to the diminished statistical uncertainty, is particularly weak and thus sensitive to systematic analysis choices; omitting it would yield $\alpha=0.81\pm0.12$, corresponding to $\Mach=2.7_{-0.3}^{+0.7}$; analogously omitting the $150\MHz$ channel in the north would give $\alpha=0.66\pm0.05$, corresponding to $\Mach=3.7_{-0.5}^{+0.7}$.

The inferred spectrum of the northern RB is robust to changes in the inspected regions, whereas the spectrum of the fainter radio bubble in the south is more sensitive to region selection.
For instance, offsetting either upstream or downstream nominal regions by up to $\pm2$ bins changes the spectrum by less than $\Delta\alpha=0.02$ in the north, whereas in the south bubble such variations modify the spectrum by up to $\Delta\alpha=0.13$ with respect to the nominal result.
Using the alternative, linear $F_\nu(\psi)$ foreground, the nominal downstream region yields $\alpha=0.588\pm0.003$ in the northern RB, corresponding to a stronger $\Mach=4.9\pm0.1$ shock, again robustly (within $\Delta\alpha=0.05$) to different choices of downstream region.
In the south, the weaker signal combined with the steep $F_\nu(\psi)$ profiles of the two low-frequency channels preclude such linear foreground removal; using only the remaining two high-frequency channels then suggests $\alpha\simeq 0.6$, 
corresponding to
$\Mach\simeq 4.6$, but this result again varies by $\Delta\alpha\simeq 0.1$ for different choices of the inspected downstream region.

Comparing the spectra of the two RBs in the bottom panel of Fig.~\ref{fig:Radio} shows that the northern bubble is nearly an order of magnitude brighter in radio than its southern counterpart.
Nevertheless, the $I_\nu(\psi)$ profiles are qualitatively similar in both hemispheres: as one approaches the bubble from outside (with decreasing $\psi>0$), the brightness initially increases slowly, but rapidly rises just inside ($\psi<0$) the edge to peak at $-8\dgr<\psi<-4\dgr$, slightly before the thermal $-14\dgr<\psi<-10\dgr$ X-ray peak.
In contrast, while the \gama-ray profile of the northern RB in Fig.~\ref{fig:GammaRay} is similar to its radio counterpart, the southern bubble appears to be superimposed on a somewhat ($10$--$40\%$) stronger \gama-ray foreground which rises with $\psi$.
Therefore, in \gama-rays, a linear foreground is needed to test for an edge in Fig.~\ref{fig:GammaRay} (top and middle panels) and to infer the RB spectrum (bottom panel).

The \gama-ray profiles $I_\nu(\psi)$ in Fig.~\ref{fig:GammaRay} indicate excess emission just downstream of the edge, in all channels and in both hemispheres.
However, a significant $>5\sigma$ shock detection is obtained only in channels 2 and 3 in the northern bubble, and in channel 1 in the south.
The figure demonstrates, for the most noisy, high-energy channel 4, linear regression fits for the $I_\nu(\psi)$ profile (dotted red curve, $\mathsf{n}_-=2$) and for its separate upstream and downstream sections (dashed, $\mathsf{n}_+=4$), indicating a shock in this channel at only the $2.9\sigma$ confidence level in the north and $1.9\sigma$ in the south.
The bottom panel shows that the implied spectra are similar in the north and south bubbles, although the uncertainties are substantial; the better constrained channels 2 and 3 suggest a somewhat brighter northern bubble.
For better statistics, the \gama-ray spectra in the bottom panel are extracted from a broader, $-8\dgr<\psi<-2\dgr$ downstream region.

\section{Summary and discussion}
\label{sec:Discussion}

Using the sharp X-ray edges we detect near the tips of both north and south RBs in eROSITA data (Fig.~\ref{fig:eROSITA}), we stack broadband data parallel to the edges, measure the jumps in radio (Fig.~\ref{fig:Radio}) and \gama-ray (Fig.~\ref{fig:GammaRay}) brightness across the edges, and extract the respective RB spectra (bottom panels of Figs.~\ref{fig:Radio} and \ref{fig:GammaRay}).
The shock Mach numbers inferred from the radio spectra depend somewhat on foreground removal method, nominally indicating $\Mach=3.5_{-0.4}^{+0.7}$ in the north and $\Mach=3.2_{-1.0}^{+1.7}$ in the south, and slightly stronger shocks in most analysis variants.
We conclude that the RB edges are strong, $3<\Mach\lesssim5$ forward shocks.

In comparison to its northern counterpart, the southern bubble is nearly an order of magnitude fainter in radio and somewhat fainter also in \gama-rays, while propagating into an upstream which is brighter in \gama-rays.
These effects render the RBs more difficult to detect in the southern hemisphere through non-thermal emission, both in radio and in \gama-rays, emphasizing the importance of stacking data along the edge.
As the inferred shocks are similarly strong in both hemispheres, the brighter RB in the north suggests a higher upstream density.
While radio emission depends on the magnetic field, and \gama-rays carry large uncertainties, the factor $4$--$5$ brighter X-ray emission from the northern RB shell (after foreground removal) with respect to its southern counterpart robustly indicates a density higher by a factor $\sim 2$.
This factor pertains to emission within several degrees of the projected edge, so folds-in any geometric asymmetry such as an outflow axis possibly tilted towards the solar system in its south.

Shock Mach numbers are inferred from low-frequency radio emission, below any spectral break, by assuming in Eq.~\eqref{eq:alpha} synchrotron emission from shock-accelerated, non-cooled cosmic-ray electrons (CREs).
Cooling is indeed negligible in the inspected frequency range, provided that the RBs are not much older than the $27U_1^{-1}(B/\muG)^{1/2}\Myr$ cooling time at $1.4\GHz$, where electrons were assumed to Compton-cool off radiation fields of total energy density $U=1 U_1\eV\cm^{-3}$ \citep[see][]{KeshetEtAl24} and gyrate in magnetic fields of mean amplitude $B$.
Alternative interpretations of the hard spectrum as compression or re-acceleration of a preexisting CRE population by a weak shock \citep[\eg][]{MouEtAl23} are disfavored because the large jump in stacked X-ray brightness (factor $>3$ even before accounting for foregrounds) indicates a strong shock, and the spectra are similar in both hemispheres despite evidence for different upstream conditions and very different bubble brightness levels.
Hadronic emission alternatives can also be ruled out because, as in the FBs, the \gama-ray spectrum observed below $1\GeV$ is too soft \citep{KeshetEtAl24}, the high-latitude ambient density is too low, and the similarity to the FB edge spectra \citep{Keshetgurwich17} in terms of both spectra and normalization is striking.

The similarities we find between the RBs and their smaller, FB counterparts, in terms of the bubble edge morphology and structure, their associated nonthermal emission \citep[\emph{cf.}][]{Keshetgurwich17,KeshetEtAl24}, and the implied high Mach numbers, indicate that the two structures share a common origin, as separate consecutive high-energy outbursts from the Galactic center, which must have been collimated roughly perpendicular to the Galactic disk \citep{MondalEtAl22}.
The polarized, large-scale radio lobes, previously attributed to the FBs \citep{CarrettiEtAl13} but later shown to be distinct from the subtle FB component of polarized radio emission \citep{Keshet25PolFB}, most likely arise from the compressed magnetic fields in the RB downstream.
A detailed analysis of the two outbursts, driving the RBs and later the FBs, is deferred to \citet{GhoshEtAl26}.

\begin{acknowledgements}
This research received funding from ISF grant No. 2126/22. \\
\end{acknowledgements}

\bibliographystyle{aa}
\bibliography{FermiBubbles}

\end{document}